\documentclass[12pt]{article}

\usepackage{amssymb}
\usepackage{amsmath}

\usepackage{makecell}

\usepackage{graphicx}

\begin{document} %

\newcommand{\bs}{\boldsymbol} 


\title{Probing anomalous $\gamma\gamma\gamma Z$ couplings
through $\gamma Z$ production in $\gamma\gamma$ collisions at the
CLIC}

\author{
S.C. \.{I}nan\thanks{Electronic address: sceminan@cumhuriyet.tr}
\\
{\small Department of Physics, Sivas Cumhuriyet University, 58140,
Sivas, Turkey}
\\
{\small and}
\\
A.V. Kisselev\thanks{Electronic address:
alexandre.kisselev@ihep.ru} \\
{\small Division of Theoretical Physics, A.A. Logunov Institute for
High Energy Physics,}
\\
{\small NRC ``Kurchatov Institute'', 142281, Protvino, Russia}}

\date{}

\maketitle

\begin{abstract}
We have estimated the sensitivity to the anomalous couplings of the
$\gamma\gamma\gamma Z$ vertex in the $\gamma\gamma\rightarrow\gamma
Z$ scattering of the Compton backscattered photons at the CLIC. Both
polarized and unpolarized collisions at the $e^+e^-$ energies 1500
GeV and 3000 GeV are addressed, and anomalous contributions to
helicity amplitudes are derived. The differential and total cross
sections are calculated. We have obtained 95\% C.L. exclusion limits
on the anomalous quartic gauge couplings (QGCs). They are compared
with corresponding bounds derived for the $\gamma\gamma\gamma Z$
couplings via $\gamma Z$ production at the LHC. The constraints on
the anomalous QGCs are one to two orders of magnitude more stringent
that at the HL-LHC. The partial-wave unitarity constraints on the
anomalous couplings are examined. It is shown that the unitarity is
not violated in the region of the anomalous QGCs studied in the
paper.
\end{abstract}


\section{Introduction} %
\label{sec:intr}

In our previous paper \cite{Inan_Kisselev:2021} we probed the
anomalous quartic gauge couplings (QGCs) in the $\gamma\gamma
\rightarrow \gamma\gamma$ process at the Compact Linear Collider
(CLIC) \cite{Braun:2008,Boland:2016}. Both the unpolarized and
polarized light-by-light scatterings were considered, and the bounds
on QGCs were obtained. The neutral anomalous quartic couplings are
of particular interest. The anomaly interactions $\gamma ZZZ$,
$\gamma\gamma ZZ$, and $\gamma\gamma\gamma Z$ at the LHC were
analyzed in \cite{Chapon:2009}-\cite{Eboli:2004}. The LHC
experimental bounds on QGCs were presented by the CMS
\cite{CMS:QGCs} and ATLAS \cite{ATLAS:QGCs} Collaborations (see also
\cite{Schoeffel:2021}). The bounds on the anomalous
$\gamma\gamma\gamma Z$ vertex can be also derived from the
constraints on the $\mathcal{B}(Z\rightarrow\gamma\gamma\gamma)$
branching ratio obtained at the LEP \cite{L3:Z_decay} and LHC
\cite{ATLAS:Z_decay}. As for $e^+e^-$ colliders, they may operate in
$e\gamma$ and $\gamma\gamma$ modes~\cite{Ginzburg:1981}. The bounds
on QGCs in $e^+e^-$, $e\gamma$ and $\gamma\gamma$ collisions were
given in \cite{Eboli:1994}-\cite{Koksal:2014}. In particular, the
limits on the quartic couplings for the vertex $\gamma\gamma\gamma
Z$ were derived in \cite{Gutierrez:2014} using LEP~2 data for the
reactions $e^+e^- \rightarrow \gamma\gamma\gamma, \gamma\gamma Z$. A
similar analysis for the exclusive $\gamma Z$ production with intact
protons at the LHC was done in \cite{Baldenegro:2017}. The search
for virtual SUSY effects in the process $\gamma\gamma \rightarrow
\gamma Z$ at high energies was presented in \cite{Gounaris:1999_1}.

\begin{figure}[htb]
\begin{center}
\includegraphics[scale=0.6]{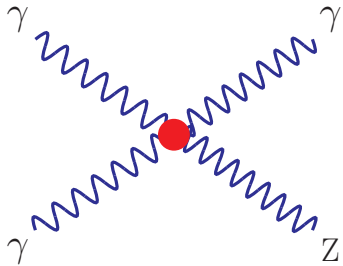}
\caption{The process $\gamma + \gamma \rightarrow \gamma + Z$.}
\label{fig:3gammasZ}
\end{center}
\end{figure}

As one can see, the anomalous $\gamma\gamma\gamma Z$ vertex urgently
needs to be examined in high energy $e^+e^-$ collisions. That is
why, in the present paper we study the process (see
Fig.~\ref{fig:3gammasZ})
\begin{equation}\label{process}
\gamma(p_1, \mu)+ \gamma(p_2, \nu) \rightarrow \gamma(p_3, \rho) +
Z(p_4, \alpha) \;,
\end{equation}
where $p_1, p_2, p_3, p_4$ are boson momenta, $\mu, \nu, \rho,
\alpha$ are boson Lorentz indices, and ingoing particles are real
polarized photons generated at the CLIC by the laser Compton
backscattering~\cite{Kramer:1994}-\cite{Telnov:1998}.
%
Our main goal is to derive bounds on anomaly couplings for the
vertex $\gamma\gamma\gamma Z$ which can be reached at the CLIC using
both polarized and unpolarized photon beams. The great potential of
the CLIC in probing new physics is well-known
\cite{Dannheim:2012}-\cite{Franceschini:2020}. Let us underline that
a physical potential of a linear high energy $e^+e^-$ collider may
be significantly enhanced, provided the polarized beams are
used~\cite{polarized_beams,CLIC_lum}.

Let $\lambda_e$ be the helicity of the initial electron beam, while
$\lambda_0$ be the helicity of the ingoing laser photon beam. In our
calculations, we will consider two sets of these helicities, with
opposite sign of $\lambda_e$,
\begin{align}\label{helicities}
(\lambda_e^{(1)}, \lambda_0^{(1)}; \lambda_e^{(2)}, \lambda_0^{(2)})
&= (0.8, 1; 0.8, 1) \;, \nonumber \\
(\lambda_e^{(1)}, \lambda_0^{(1)}; \lambda_e^{(2)}, \lambda_0^{(2)})
&= (-0.8, 1; -0.8, 1) \;,
\end{align}
where the superscripts 1 and 2 enumerate the beams. We will work in
the effective field theory framework. Previously effective
Lagrangians were used in \cite{Stohr:1994}-\cite{Novotny:1995} for
examining the $\gamma\gamma\gamma Z$ interaction in the $Z
\rightarrow \gamma\gamma\gamma$ decay, as well as in
\cite{Baillargeon:1996}, \cite{Gutierrez:2014}, and
\cite{Baldenegro:2017}. Anomalous quartic gauge couplings (QGCs) are
induced at the dimension-six level already. However, they are not
independent of anomalous trilinear gauge couplings. That is why, in
our paper, we study anomalous QGCs which enter the effective
Lagrangian at dimension-eight without contributing to anomalous
trilinear gauge interactions.

The paper is organized as follows. In the next section, the effective
Lagrangian is described, and Feynman rules for the anomalous
$\gamma\gamma\gamma Z$ vertex are presented. The helicity amplitudes
are studied in Sec.~\ref{sec:amplitudes}. In
Sec.~\ref{sec:numerical_results}, both differential and total cross
sections for the process \eqref{process} are calculated, and bounds
on the QGCs are given. In Sec.~\ref{sec:unit_const}, unitarity
constraints on anomalous quartic couplings are obtained. In
Appendix~A, polarization tensors for the vertex $\gamma\gamma\gamma
Z$ are listed. The explicit expressions for the anomalous
contributions to the helicity amplitudes are given in Appendix~B.
Some formulas for Wigner's $d$-function are collected in Appendix~C.
Finally, in Sec.~\ref{sec:concl}, we summarize our results and give
conclusions.

\section{Effective Lagrangian} %
\label{sec:Lagrangian}

It is appropriate to describe the anomalous $\gamma\gamma\gamma Z$
interaction by means of an effective Lagrangian. Given parity is
conserved and gauge invariance is valid, there are only two
independent operators with dimension 8. Following
\cite{Stohr:1994,Horejsi:1994}, we take the Lagrangian
\begin{equation}\label{Lagrangian}
\mathcal{L}_{\gamma\gamma\gamma Z} = g_1 O_1 + g_2 O_2 \;,
\end{equation}
with the operators
\begin{equation}\label{operators_G-R}
O_1 = F^{\rho\mu} F^{\alpha\nu} \partial_\rho F_{\mu\nu} Z_\alpha
\;, \quad O_2 = F^{\rho\mu} F^\nu_\mu \partial_\rho F_{\alpha\nu}
Z^\alpha \;,
\end{equation}
where  $F_{\mu\nu} = \partial_\mu A_\nu - \partial_\nu A_\mu$. The
operators $O_{1,2}$ arise from a $SU(2)\times U(1)_Y$ effective
Lagrangian with two operators like
$B_{\mu\nu}B^{\nu\nu}B_{\rho\sigma}B^{\rho\varrho}$, four operators
like $W_{\mu\nu}W^{\nu\nu}W_{\rho\sigma}W^{\rho\varrho}$, and four
operators like $B_{\mu\nu}B^{\nu\nu}W_{\rho\sigma}W^{\rho\varrho}$,
where $W_\mu$ and $B_\mu$ are the $SU(2)$ and hypercharge gauge
fields, respectively \cite{Fichet:2014}. We consider only CP
conserving operators hence the dual field strength tensors
$\tilde{W}_{\mu\nu}$ and $\tilde{B}_{\mu\nu}$ are not used. The
coupling $g_{1,2}$ are linear combinations of ten coefficients of
dimension-eight operators mentioned above. Note that certain
combinations of these coefficients must obey so-called positivity
constraints \cite{Remmen:2019}-\cite{Bi:2019}.

As one can see, this Lagrangian contains no derivatives of the $Z$
boson field (correspondingly, no $p_4$ in the momentum space), that
simplifies a derivation of Feynman rules for the $\gamma\gamma\gamma
Z$ vertex.

Some authors use the Lagrangian \cite{Novotny:1995}
\begin{equation}\label{Lagrangian_N}
\mathcal{L}^{(\mathrm{N})}_{\gamma\gamma\gamma Z} = G_1 \bar{O}_1 +
G_2 \bar{O}_2 \;,
\end{equation}
with the operators
\begin{equation}\label{operators_B}
\bar{O}_1 =  F^{\mu\nu}F_{\mu\nu} F^{\rho\sigma}Z_{\rho\sigma} \;,
\quad \bar{O}_2 = F^{\mu\nu} F_{\nu\rho}F^{\rho\sigma}Z_{\sigma\mu}
\;,
\end{equation}
where $Z_{\mu\nu} = \partial_\mu Z_\nu - \partial_\nu Z_\mu$, or the
Lagrangian \cite{Baldenegro:2017}
\begin{equation}\label{Lagrangian_B}
\mathcal{L}^{(\mathrm{B})}_{\gamma\gamma\gamma Z} = \zeta O +
\tilde{\zeta} \tilde{O} \;,
\end{equation}
with the operators
\begin{equation}\label{operators_B}
O = F^{\mu\nu}F_{\mu\nu} F^{\rho\sigma}Z_{\rho\sigma} \;, \quad
\tilde{O} = F^{\mu\nu}
\tilde{F}_{\mu\nu}F^{\rho\sigma}\tilde{Z}_{\rho\sigma} \;,
\end{equation}
where $\tilde{F}_{\mu\nu} =
(1/2)\,\varepsilon^{\mu\nu\rho\sigma}F_{\varrho\sigma}$, and
$\tilde{Z}_{\mu\nu} =
(1/2)\,\varepsilon^{\mu\nu\rho\sigma}Z_{\varrho\sigma}$.

Using integration by parts and equations of motion, one can easily
obtain the following relations between two bases for the effective
Lagrangian \cite{Novotny:1995},
\begin{equation}\label{operators_N_vs_G-R}
\bar{O}_1 = -8O_1 \;, \quad \bar{O}_2 = 2(O_2 - O_1) \;.
\end{equation}
We also have the relations \cite{Baldenegro:2017}
\begin{equation}\label{operators_G-R_vs_B}
O = \bar{O}_1 = -8 O_1 \;, \quad \tilde{O} = 4\bar{O}_2 - 2\bar{O}_1
= 8(O_2 + O_1) \;.
\end{equation}
The above listed equations enable us to relate anomalous coupling in
eqs.~\eqref{Lagrangian}, \eqref{Lagrangian_N}, and
\eqref{Lagrangian_B}. In particular, we find
\begin{equation}\label{couplings_G-R_B}
g_1 = 8(\tilde{\zeta} - \zeta) \;, \quad g_2 = 8 \tilde{\zeta} \;.
\end{equation}

The Feynman rules for the effective anomalous vertex, resulting from
the Lagrangian \eqref{Lagrangian}, are given by \cite{Horejsi:1994}
\begin{align}\label{Feynman_rules}
P^{\mu\nu\rho\alpha} = \mathcal{P}&\{ g_1[(p_1\cdot p_2)(p_2\cdot
p_3) g^{\mu\nu}g^{\rho\alpha} - (p_1\cdot p_3) p_2^\mu p_1^\nu
g^{\rho\alpha}
\nonumber \\
&- (p_1\cdot p_3)p_1^\nu p_2^\alpha g^{\mu\rho} + p_2^\mu p_1^\nu
p_1^ \rho p_3^\alpha ]
\nonumber \\
&+ g_2[ -(p_1\cdot p_2)(p_1\cdot p_3) g^{\mu\alpha}g^{\nu\rho} +
(p_2\cdot p_3) p_1^\nu p_1^\alpha g^{\mu\rho}
\nonumber \\
&- (p_2\cdot p_3) p_1^\nu p_1^\rho g^{\mu\alpha} + (p_2\cdot p_3)
p_1^\nu p_2^\alpha g^{\mu\rho} + 2(p_2\cdot p_3) p_2^\mu p_1^\rho
g^{\nu\alpha}
\nonumber \\
&- (p_1\cdot p_3) p_2^\rho p_1^\alpha g^{\mu\nu} + p_3^\mu p_1^\nu
p_2^\rho p_1^\alpha ] \} \;,
\end{align}
where $\mathcal{P}$ denotes possible permutations $(p_1,\mu)
\leftrightarrow (p_2,\nu) \leftrightarrow (p_3, \rho)$, and all
momenta in the $\gamma\gamma\gamma Z$ vertex are assumed to be
incoming ones. Correspondingly, the polarization tensor is equal to
\begin{equation}\label{polarization_tensor}
P_{\mu\nu\rho\alpha}(p_1, p_2, p_3) = g_1 \sum_{i=1}^4
P_{\mu\nu\rho\alpha}^{(1,i)}(p_1, p_2, p_3) + g_2 \sum_{i=1}^7
P_{\mu\nu\rho\alpha}^{(2,i)}(p_1, p_2, p_3) \;.
\end{equation}
Electromagnetic gauge invariance results in equations $p_1^\mu
P_{\mu\nu\rho\alpha} = p_2^\nu P_{\mu\nu\rho\alpha} = p_3^\rho
P_{\mu\nu\rho\alpha} = 0$. Note that terms proportional to $p_1^\mu,
p_2^\nu, p_3^\rho$ are omitted in \eqref{Feynman_rules}, since they
do not contribute to the matrix element, see
eq.~\eqref{matrix_element} below. Explicit expressions for the
tensors $P_{\mu\nu\rho\alpha}^{(1,i)}$ and
$P_{\mu\nu\rho\alpha}^{(2,i)}$ are presented in Appendix~A. To
calculate helicity amplitudes for the process \eqref{process}, one
has to make the replacement $p_3 \rightarrow -p_3$ in the Feynman
rules for the $\gamma\gamma\gamma Z$ vertex given by
eqs.~\eqref{Feynman_rules}, \eqref{polarization_tensor}, and
\eqref{P1.1}-\eqref{P2.7}.

\section{Helicity amplitudes} %
\label{sec:amplitudes}

We work in the c.m.s. of the colliding real photons, $\vec{p}_1 +
\vec{p}_2 = 0$, where the momenta are given by
\begin{align}\label{momenta}
p_1^\mu &=(p,0,0,p) \;,
\nonumber \\
p_2^\mu &=(p,0,0,-p) \;,
\nonumber \\
p_3^\mu &=(k, 0, k\sin\theta, k\cos\theta) \;,
\nonumber \\
p_4^\mu &=(E, 0, -k\sin\theta, -k\cos\theta) \;.
\end{align}
Here $E = \sqrt{k^2 + m_Z^2}$, with $m_Z$ being the mass of the $Z$
boson. The Mandelstam variables of the process \eqref{process} are
\begin{align}\label{Mandelstam_var}
s &= (p_1 + p_2)^2 = 4p^2 \;,
\nonumber  \\
t &= (p_1 - p_3)^2 = -2pk (1 - \cos\theta) \;,
\nonumber  \\
u &= (p_2 - p_3)^2 = -2pk (1 + \cos\theta) \;,
\end{align}
where $\theta$ is a scattering angle in the c.m.s. Note that $s + t
+ u = m_Z^2$.

In the chosen system the polarization vectors are equal to
\begin{align}\label{pol_vectors_initial}
\varepsilon_\mu^+(p_1) &= \varepsilon_\mu^-(p_2) =
\frac{1}{\sqrt{2}}(0, 1, i, 0) \;,
\nonumber \\
\varepsilon_\mu^-(p_1) &= \varepsilon_\mu^+(p_2) =
\frac{1}{\sqrt{2}}(0, 1, -i, 0) \;,
\nonumber \\
\varepsilon_\mu^+(p_3) &= \varepsilon_\mu^-(p_4) =
\frac{1}{\sqrt{2}}(0, 1, i\cos \theta, -i\sin\theta) \;,
\nonumber \\
\varepsilon_\mu^-(p_3) &= \varepsilon_\mu^+(p_4) =
\frac{1}{\sqrt{2}}(0, 1, -i\cos\theta,
i\sin\theta) \;, \nonumber \\
\varepsilon_\mu^0(p_4) &= \frac{1}{m_Z}(k, 0, -E\sin\theta, -
E\cos\theta) \;.
\end{align}
They obey the orthogonality condition $\varepsilon_\mu^\lambda (k)
k^\mu = 0$. Correspondingly, we get the helicity vectors of the
final photon and $Z$ boson,
\begin{align}\label{pol_vectors_final}
\varepsilon_\mu^{*+}(p_3) &= \varepsilon_\mu^{*-}(p_4) =
\frac{1}{\sqrt{2}}(0, 1, -i\cos \theta, i\sin\theta) \;,
\nonumber \\
\varepsilon_\mu^{*-}(p_3) &= \varepsilon_\mu^{*+}(p_4) =
\frac{1}{\sqrt{2}}(0, 1, i\cos\theta, -i\sin\theta) \;,
\nonumber \\
\varepsilon_\mu^{*0}(p_4) &= \varepsilon_\mu^0(p_4) =
\frac{1}{m_Z}(k, 0, -E\sin\theta, - E\cos\theta)  \;.
\end{align}

The matrix element of the process \eqref{process} with the definite
helicities of the incoming and outgoing bosons can be written as
\begin{equation}\label{matrix_element}
M_{\lambda_1\lambda_2\lambda_3\lambda_4}(p_1, p_2, p_3) =
P_{\mu\nu\rho\alpha}(p_1, p_2, p_3)
\,\varepsilon_\mu^{\lambda_1}(p_1)\varepsilon_\nu^{\lambda_2}(p_2)
\varepsilon_\rho^{*\lambda_3}(p_3)\varepsilon_\alpha^{*\lambda_4}(p_4)
\;,
\end{equation}
where the polarization tensor $P_{\mu\nu\rho\alpha}$ is given by
eq.~\eqref{polarization_tensor}. We have calculated the anomalous
helicity amplitudes, and present their explicit expressions in
Appendix~B. Using these expressions, we obtain the unpolarized
amplitude squared
\begin{equation}\label{M2}
\sum_{\lambda_1\ldots\lambda_4}|M_{\lambda_1\lambda_2\lambda_3\lambda_4}|^2
= \frac{1}{4} [g_1^2(3A + 2B) - 4g_1g_2(A + B) + 4g_2^2(A + B)] \;,
\end{equation}
where
\begin{equation}\label{A_B}
A = s^2t^2 + t^2u^2 + u^2s^2 \;, \quad B = stu\,m_Z^2 \;.
\end{equation}
With a help of relations \eqref{couplings_G-R_B}, we get from
\eqref{M2} the differential cross section
\begin{align}\label{M2_B}
\frac{d\sigma_{\gamma\gamma\rightarrow\gamma Z}}{d\Omega} &=
\frac{\beta}{64\pi^2 s} \frac{1}{4}
\sum_{\lambda_1\ldots\lambda_4}|M_{\lambda_1\lambda_2\lambda_3\lambda_4}|^2
\nonumber \\
&= \frac{\beta}{16\pi^2 s}\,[(3\zeta^2 + 3\tilde{\zeta}^2 - 2\zeta
\tilde{\zeta})A + 2(\zeta^2 + \tilde{\zeta}^2)B] \;,
\end{align}
where $\beta = 1 - m_Z^2/s$, in a full agreement with eq.~(2.3) in
\cite{Baldenegro:2017}.

To estimate a SM background, we take analytical expressions for the
SM helicity amplitudes from Appendix~A in \cite{Gounaris:1999_1}.
Both $W$ boson loops \cite{Bailargeon:1991,Jikia:1994} and charged
fermion loops \cite{Bailargeon:1991,Bij:1989} contribute to these
amplitudes. As shown in \cite{Gounaris:1999_1}, for $s > (250 \
\mathrm{GeV})^2$ the dominant SM amplitudes
$A_{\lambda_1\lambda_2\lambda_3\lambda_4}$ are the $W$-loop non-flip
amplitudes $A^W_{++++}(s,t,u)$ and $A^W_{+-+-}(s,t,u) =
A^W_{+--+}(s,u,t)$. Almost negligible are $A^W_{+++0}(s,t,u)$ and
$A^W_{+-+0}(s,t,u) = A^W_{+--0}(s,u,t)$. The rest are even smaller.
The fermion-loop amplitudes are comparable only to very small
$W$-loop amplitudes \cite{Gounaris:1999_1}. Similar properties of
the SM helicity amplitudes are also valid for the process
$\gamma\gamma\rightarrow\gamma\gamma$ \cite{Gounaris:1999_2}.

Another possible background comes from the SM process $\gamma\gamma
\rightarrow \gamma l^+l^-$ where the invariant mass of the lepton
pair, $m_{l^+l^-}$, is close to the Z boson mass $m_Z$. We have
obtained the cross section of the process to be of order 10$^{-3}$
fb for $|m_{l^+l^-} - m_Z| < 10$ GeV. So, this background can be
safely ignored.

\section{Numerical results} %
\label{sec:numerical_results}

The differential cross section of the process
$\gamma\gamma\rightarrow\gamma Z$ depends on spectra of the Compton
backscattered (CB) photons $f_{\gamma/e}(x_i)$, their helicities
$\xi(E_\gamma^{(i)}, \lambda_0)$ ($i = 1,2$), and helicity
amplitudes \cite{Inan_Kisselev:2021,Sahin:2009},
\begin{align}\label{diff_cs}
\frac{d\sigma}{d\cos \theta} &= \frac{\beta}{128\pi s}
\int\limits_{x_{1 \min}}^{x_{\max}} \!\!\frac{dx_1}{x_1}
\,f_{\gamma/e}(x_1) \int\limits_{x_{2 \min}}^{x_{\max}}
\!\!\frac{dx_2}{x_2} \,f_{\gamma/e}(x_2)
\nonumber \\
&\times \bigg\{ \left[ 1 + \xi \left( E_\gamma^{(1)},\lambda_0^{(1)}
\right) \right] \left[ 1 + \xi \left( E_\gamma^{(2)},\lambda_0^{(2)}
\right) \right] \sum_{\lambda_3 \lambda_4}|M_{++\lambda_3
\lambda_4}|^2
\nonumber \\
&\quad + \left[ 1 + \xi \left( E_\gamma^{(1)},\lambda_0^{(1)}
\right) \right] \left[ 1 - \xi \left( E_\gamma^{(2)},\lambda_0^{(2)}
\right) \right] \sum_{\lambda_3 \lambda_4} |M_{+-\lambda_3
\lambda_4}|^2
\nonumber \\
&\quad + \left[ 1 - \xi \left( E_\gamma^{(1)},\lambda_0^{(1)}
\right) \right] \left[ 1 + \xi \left( E_\gamma^{(2)},\lambda_0^{(2)}
\right) \right] \sum_{\lambda_3 \lambda_4}|M_{-+\lambda_3
\lambda_4}|^2
\nonumber \\
&\quad + \left[ 1 - \xi \left( E_\gamma^{(1)},\lambda_0^{(1)}
\right) \right] \left[ 1 - \xi \left( E_\gamma^{(2)},\lambda_0^{(2)}
\right) \right] \sum_{\lambda_3 \lambda_4} |M_{--\lambda_3
\lambda_4}|^2 \bigg\} ,
\end{align}
where $\lambda_3 = +, -$, $\lambda_4 = +, -, 0$, $x_1 =
E_{\gamma}^{(1)}/E_e$ and $x_2 = E_{\gamma}^{(2)}/E_e$ are the
energy fractions of the CB photon beams, $x_{1 \min} =
p_\bot^2/E_e^2$, $x_{2 \min} = p_\bot^2/(x_{1} E_e^2)$, and
$p_{\bot}$ is the transverse momentum of the outgoing particles.
Note that $\sqrt{s x_1 x_2}$ is the invariant energy of the
backscattered photons. The explicit expressions for
$f_{\gamma/e}(x_i)$ and $\xi(E_\gamma^{(i)}, \lambda_0)$ can be
found in \cite{Inan_Kisselev:2021}.

\begin{figure}[htb]
\begin{center}
\hspace*{-0.4cm}
\includegraphics[scale=0.6]{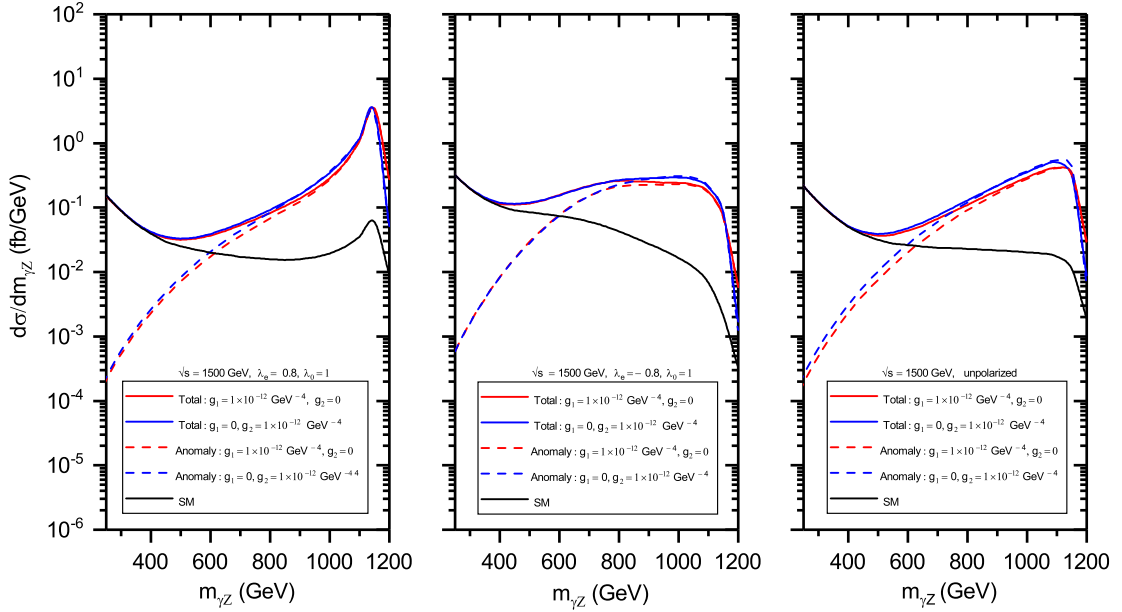}
\caption{The differential cross sections for the process
$\gamma\gamma\rightarrow\gamma Z$ as functions of the invariant mass
of the outgoing bosons for the CLIC energy $\sqrt{s} = 1500$ GeV.
The left, middle and right panels correspond to the electron beam
helicities $\lambda_e = 0.8, -0.8$, and 0, respectively. On each
plot the curves denote (from the top downwards) the differential
cross sections for the couplings $g_1 = 10^{-12} \mathrm{\
GeV}^{-4}$, $g_2 = 0$, and $g_1 = 0$, $g_2 = 10^{-12} \mathrm{\
GeV}^{-4}$, the anomalous contributions for the same values of
couplings, the SM cross section.}
\label{fig:WTDE750}
\end{center}
\end{figure}
%
\begin{figure}[htb]
\begin{center}
\hspace*{-0.4cm}
\includegraphics[scale=0.6]{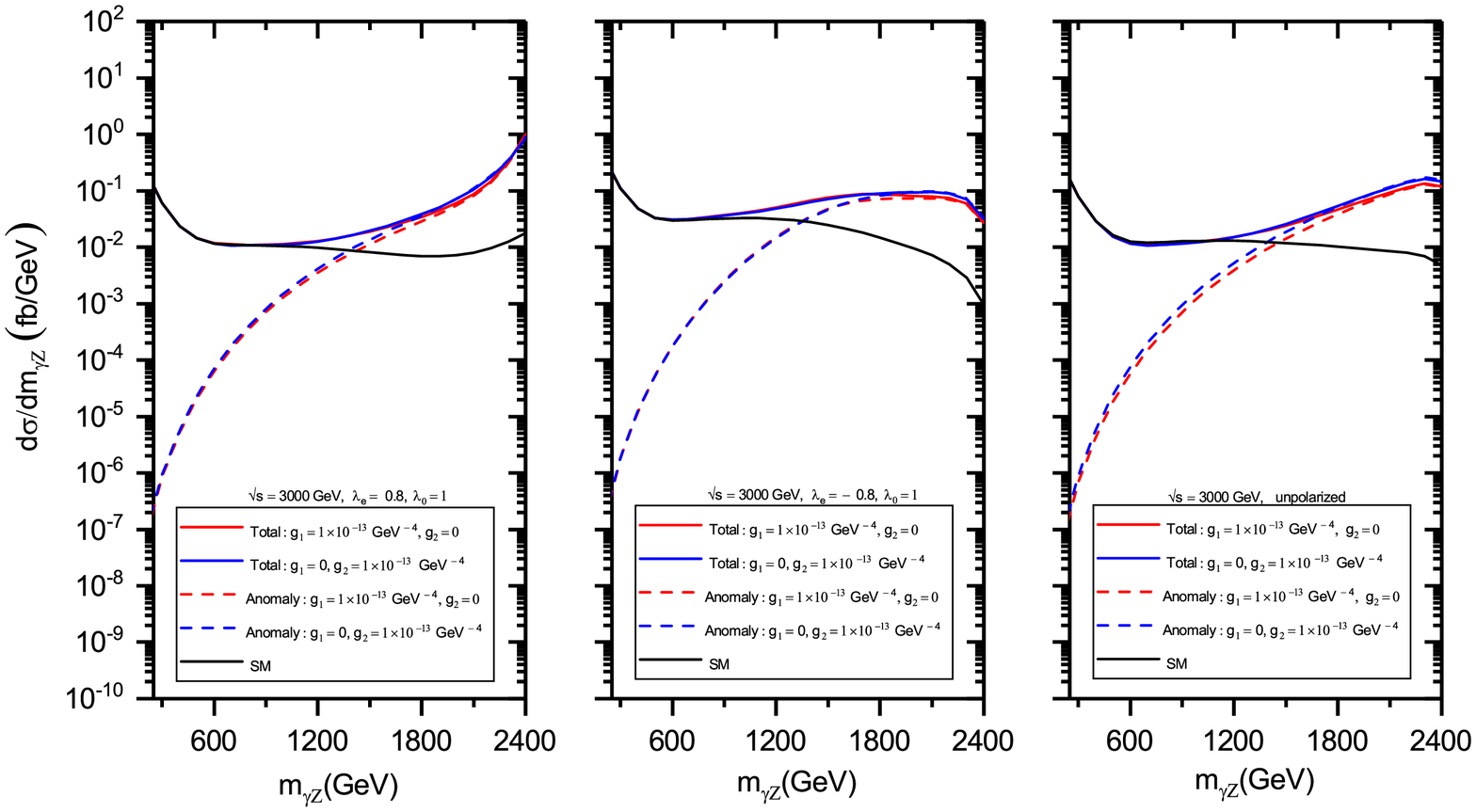}
\caption{The same as in Fig.~\ref{fig:WTDE750}, but for the $e^+e^-$
collider energy $\sqrt{s} = 3000$ GeV, and coupling sets $g_1 =
10^{-13} \mathrm{\ GeV}^{-4}$, $g_2 = 0$, and $g_1 = 0$, $g_2 =
10^{-13} \mathrm{\ GeV}^{-4}$.} \label{fig:WTDE1500F}
\end{center}
\end{figure}

\begin{figure}[htb]
\begin{center}
\includegraphics[scale=0.6]{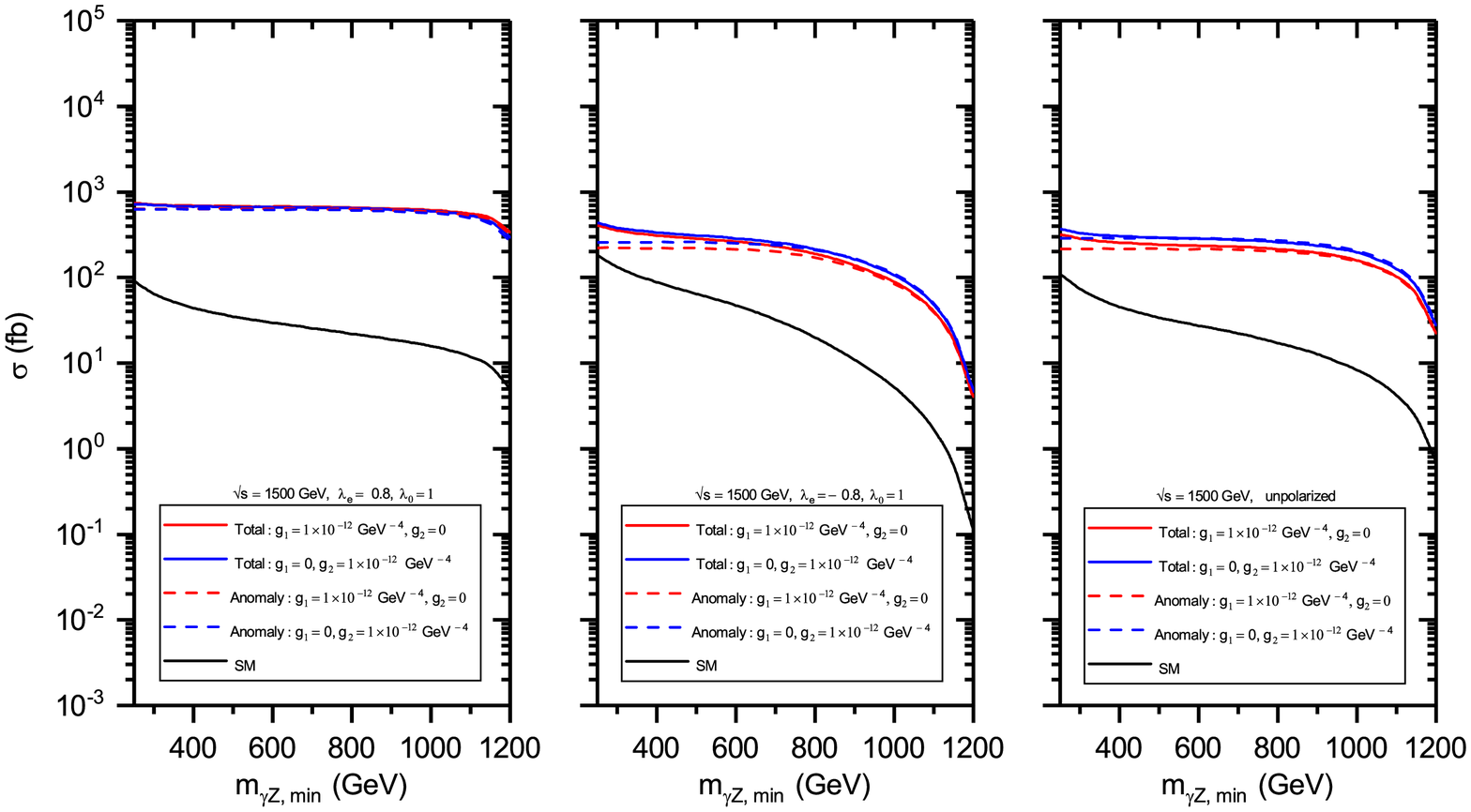}
\caption{The total cross sections for the process
$\gamma\gamma\rightarrow\gamma Z$ as functions of the minimal
invariant mass of the outgoing bosons for the $e^+e^-$ collider
energy $\sqrt{s} = 1500$ GeV. The left, middle and right panels
correspond to the electron beam helicities $\lambda_e = 0.8, -0.8$,
and 0, respectively. On each plot the curves denote (from the top
downwards) the total cross sections for the couplings $g_1 =
10^{-12} \mathrm{\ GeV}^{-4}$, $g_2 = 0$, and $g_1 = 0$, $g_2 =
10^{-12} \mathrm{\ GeV}^{-4}$, the anomalous contributions for the
same values of couplings, the SM cross section.}
\label{fig:WCUTE750}
\end{center}
\end{figure}

\begin{figure}[htb]
\begin{center}
\includegraphics[scale=0.6]{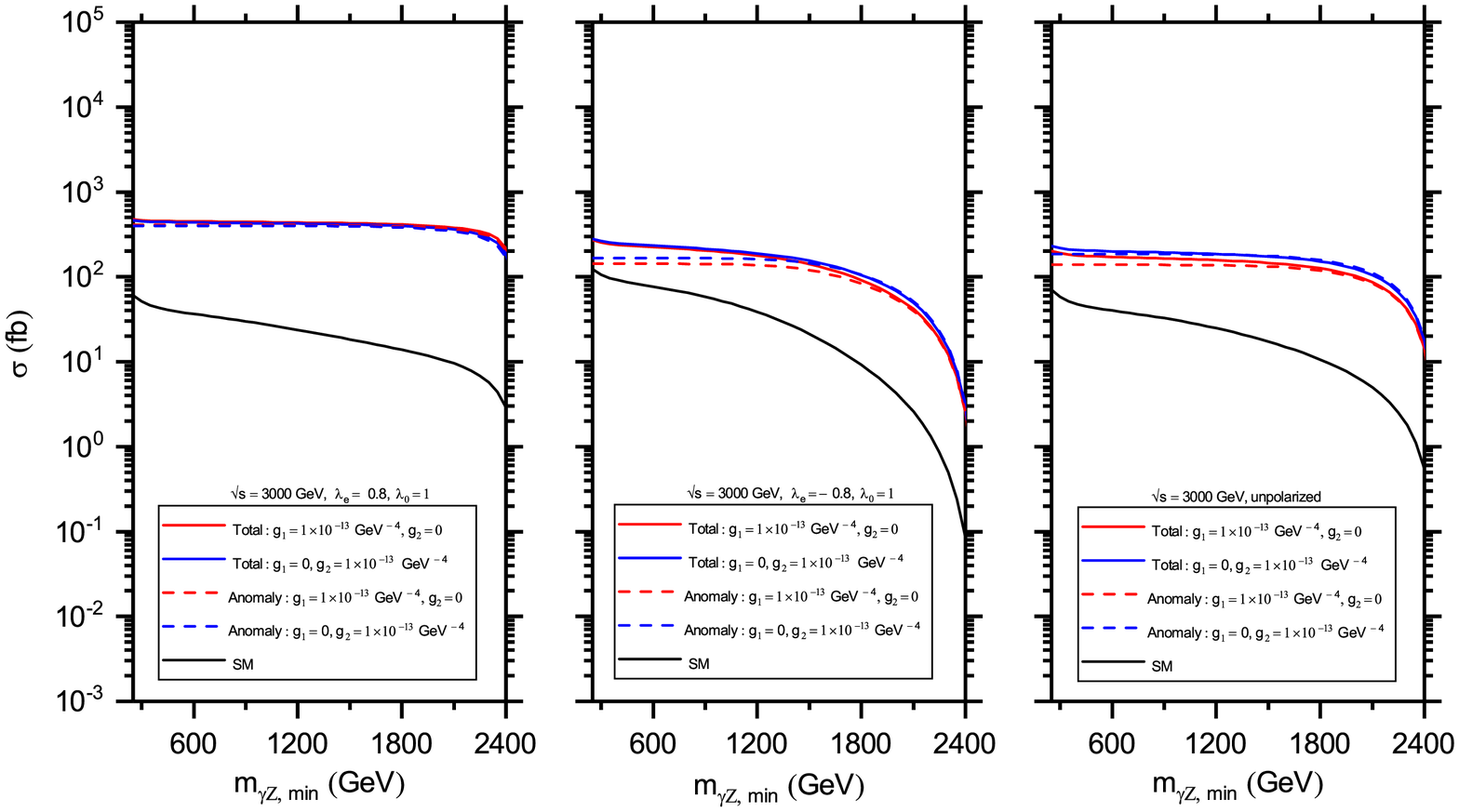}
\caption{The same as in Fig.~\ref{fig:WCUTE750}, but for the
$e^+e^-$ collider energy $\sqrt{s} = 3000$ GeV, and coupling sets
$g_1 = 10^{-13} \mathrm{\ GeV}^{-4}$, $g_2 = 0$, and $g_1 = 0$, $g_2
= 10^{-13} \mathrm{\ GeV}^{-4}$.}
\label{fig:WCUTE1500F}
\end{center}
\end{figure}

The differential cross sections are shown in
Figs.~\ref{fig:WTDE750}, \ref{fig:WTDE1500F} as functions of the
invariant mass of the $\gamma Z$ system. We have imposed the cut on
the rapidity of the final bosons, $|\eta| < 2.5$, and considered the
region $m_{\gamma Z} > 250$ GeV. As one can see, the anomalous cross
sections dominate the SM one for $m_{\gamma Z} > 600$ GeV. The
effect is more pronounced for the collision energy $\sqrt{s} = 3000$
GeV, especially as $m_{\gamma Z}$ grows. Note that for $\sqrt{s} =
3000$ GeV the differential cross sections depend weakly on electron
beam helicity $\lambda_e$. In Figs.~\ref{fig:WCUTE750},
\ref{fig:WCUTE1500F} the total cross sections are presented
depending on $m_{\gamma Z, \mathrm{min}}$, minimal invariant mass of
two outgoing bosons. The anomalous contribution dominates both the
interference one and SM cross section. The ratio of the total cross
section to the SM one grows with an increase of $m_{\gamma Z}$,
being more than one order of magnitude at large $m_{\gamma Z}$.

%

The knowledge of the total cross sections and planned CLIC
integrated luminosities \cite{CLIC_lum} enables us to calculate the
exclusion regions for the QGCs. In our study we consider leptonic
(electrons and muons) decays of the $Z$ boson. Let $s(b)$ be the
total number of signal (background) events, and $\delta$ the
percentage systematic error. The number of events is defined as
$\sigma \times L \times \mathcal{B}(Z \rightarrow e, \mu)$. The
exclusion significance is given by \cite{Zhang:2020}
\begin{equation}\label{S_excl}
S_{\mathrm{excl}} = \sqrt{ 2\left[ s - b \ln \left( \frac{b + s +
x}{2b} \right) - \frac{1}{\delta^2}\ln \left( \frac{b - s + x}{2b}
\right) - (b + s -x) \left( 1 + \frac{1}{\delta^2 b} \right) \right]
} \;,
\end{equation}
where
\begin{equation}\label{x}
x = \sqrt{(s+b)^2 - 4\delta^2 s b^2/(1 + \delta^2 b)} \;.
\end{equation}
We define the regions $S_{\mathrm{excl}} \leqslant 1.645$ as a
regions that can be excluded at the 95\% C.L. in the process
$\gamma\gamma \rightarrow \gamma Z$ at the CLIC. To reduce the SM
background, we impose the cut $m_{\gamma Z} > 1000$ GeV, in addition
to the bound $|\eta| < 2.5$. The expected integrated luminosity at
the CLIC can be found, for instance, in \cite{CLIC_lum}.

\begin{figure}[htb]
\begin{center}
\includegraphics[scale=0.6]{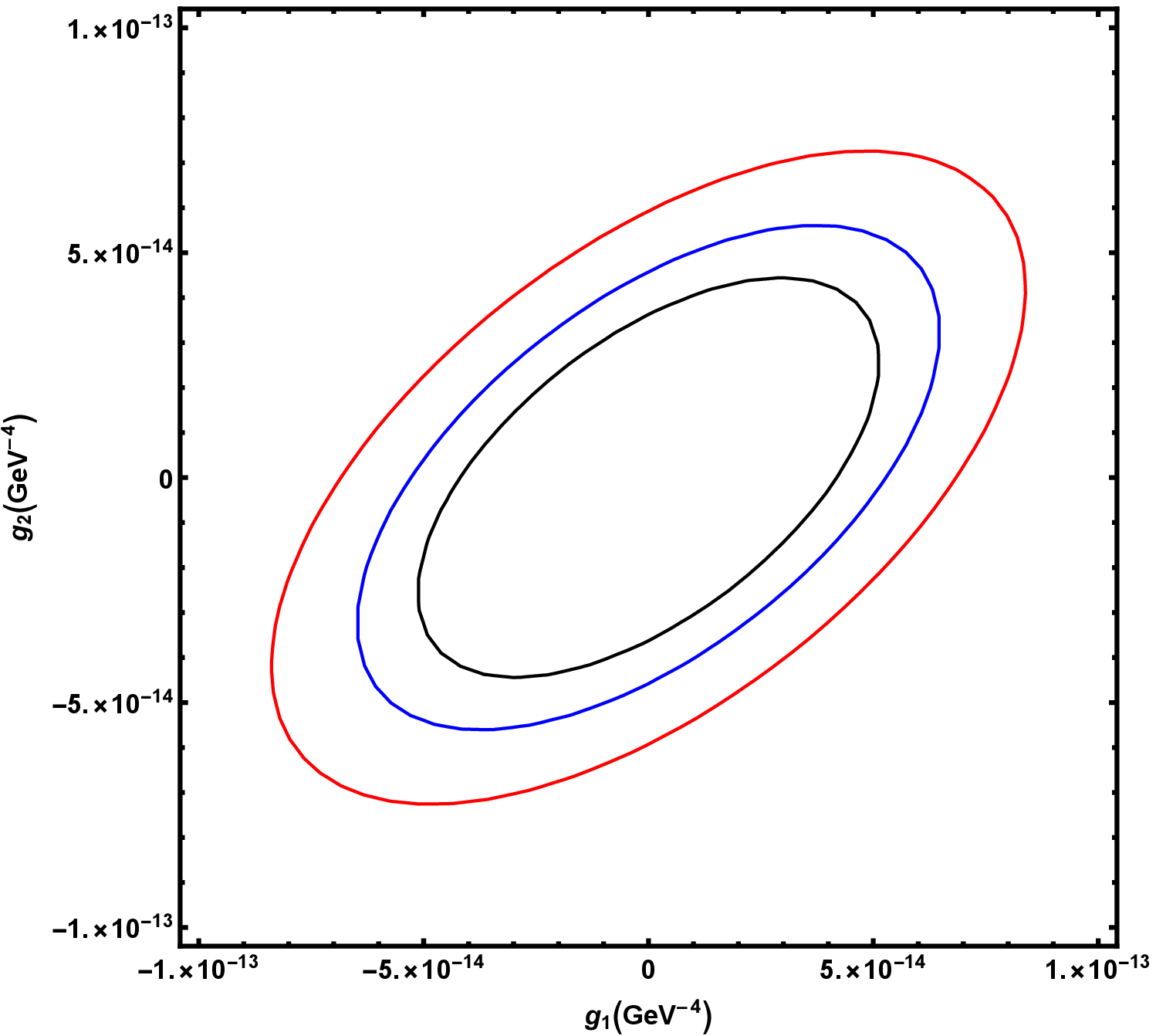}
\caption{The 95\% C.L. exclusion regions for the couplings $g_1,
g_2$ in the unpolarized reaction $\gamma\gamma \rightarrow \gamma Z$
at the CLIC with the systematic errors $\delta = 0\%$ (black
ellipse), $\delta = 5\%$ (blue ellipse), and $\delta = 10\%$ (red
ellipse). The inner regions of the ellipses are inaccessible. The
collision energy is $\sqrt{s} = 1500$ GeV, the integrated luminosity
is $L = 2500$ fb$^{-1}$. The cut on the outgoing photon invariant
mass $m_{\gamma\gamma} > 1000$ GeV was imposed.}
\label{fig:excl_750}
\end{center}
\end{figure}

\begin{figure}[htb]
\begin{center}
\includegraphics[scale=0.6]{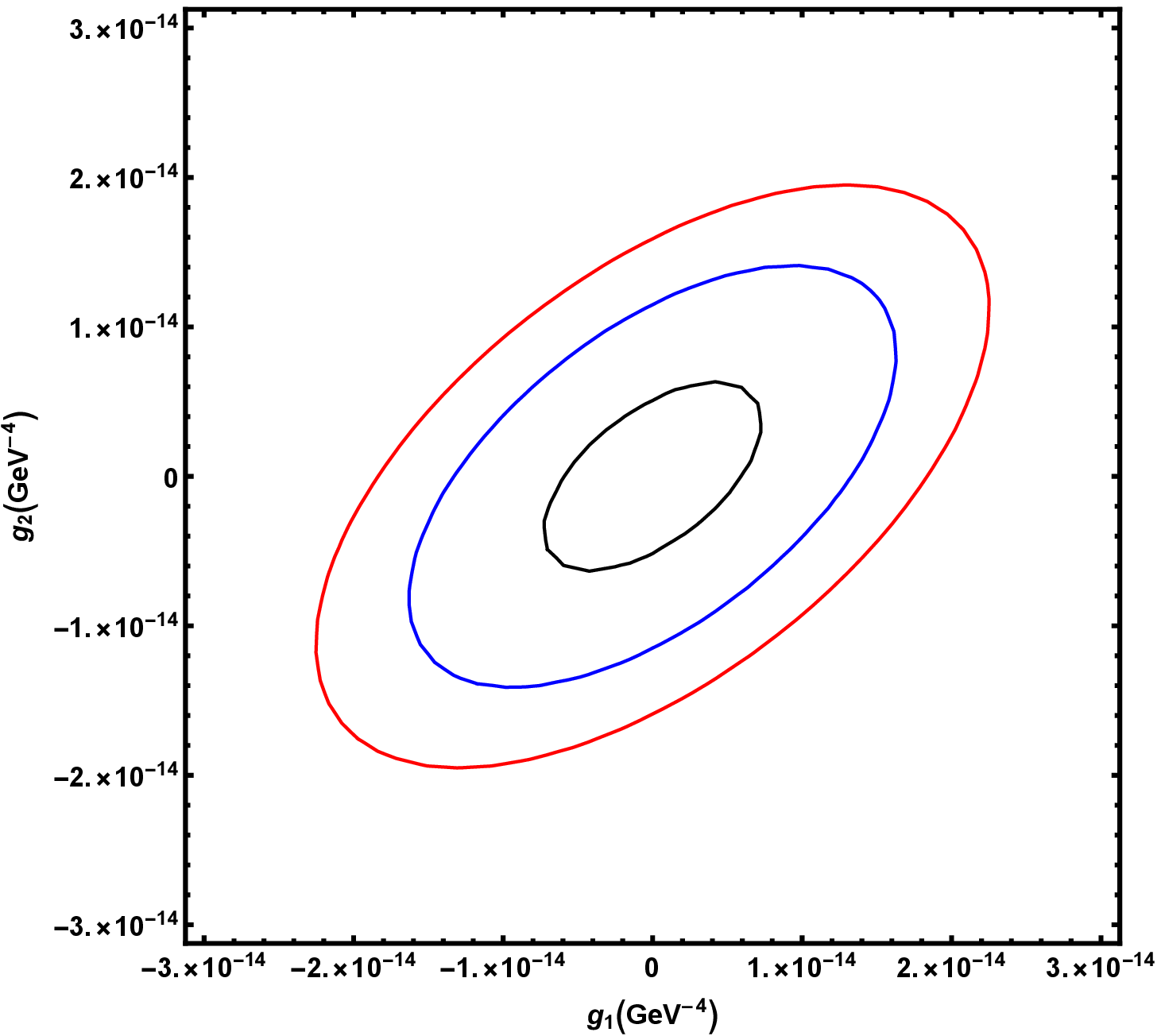}
\caption{The same as in Fig.~\ref{fig:excl_750}, but for $\sqrt{s} =
3000$ GeV and $L=5000$ fb$^{-1}$.}
\label{fig:excl_1500}
\end{center}
\end{figure}

It is worth considering the unpolarized case first. One can obtain
from eq.~\eqref{M2} that the anomalous contribution to the
unpolarized total cross section is proportional to the coupling
combination $3g_1^2 - 4g_1 g_2 + 4g_2^2$, provided terms
proportional to $m_Z^2/s \ll 1$ are neglected in it. In such a case,
the exclusion regions are ellipses in the plane $(g_1 - g_2)$
rotated clockwise through the angle $0.5 \arctan 8 \simeq
41.4^\circ$ around the origin. It is clear that our process is
slightly more sensitive to the coupling $g_2$ rather than to $g_1$.
Our 95\% C.L. exclusion regions for anomalous QGCs for the
unpolarized process $\gamma\gamma\rightarrow\gamma Z$ at the CLIC
are shown in Figs.~\ref{fig:excl_750}, \ref{fig:excl_1500}. The
results are presented for $\delta = 0$, $\delta = 5\%$, and $\delta
= 10\%$.

%

In Tabs.~\ref{tab:excl_750}, \ref{tab:excl_1500} we show the
exclusion bounds on the couplings $g_1$ and $g_2$ for three values
of the electron beam helicity $\lambda_e$ and corresponding
integrated luminosity $L$. Let us underline that this time we did
not neglect the terms proportional to $m_Z^2$, both for unpolarized
and for polarized reactions. As one can see, the best bound on the
couplings $g_{1,2}$ is approximately $5 \times 10^{-15}$ GeV$^{-4}$
for the $e^+e^-$ energy $\sqrt{s} = 3000$ GeV and electron beam
helicity $\lambda_e = 0.8$.

\begin{table}[h]
    \centering \caption{The 95\% C.L. exclusion limits on the anomalous
        quartic couplings $g_1$ and $g_2$  for the collision energy
        $\sqrt{s} = 1500$ GeV, and the cut $m_{\gamma Z} > 1000$ GeV.
        \bigskip} \label{tab:excl_750}
    \begin{tabular}{||c||c|c|c|c||}
        \hline
        $\lambda_e$ & & 0 & $-0.8$ & 0.8 \\
        \hline
        \quad $L$, fb$^{-1}$ & & 2500 & 2000 & 500 \\
        \hline
        \makecell{$|g_1|$, GeV$^{-4}$ \\ $(g_2=0)$} & \makecell{$\delta=0\%$ \\  $\delta=5\%$ \\ $\ \, \delta=10\%$ } &
        $\makecell{ 4.19\times10^{-14} \\ 5.32\times10^{-14} \\ 6.81\times10^{-14}}$ &
        $\makecell{ 6.25\times10^{-14} \\ 7.91\times10^{-14} \\ 1.02\times10^{-13}}$ &
        $\makecell{ 4.42\times10^{-14} \\ 5.38\times10^{-14} \\ 6.78\times10^{-14}}$ \\
        \hline
        \makecell{$|g_2|$, GeV$^{-4}$ \\  $(g_1=0)$} & \makecell{$\delta=0\%$ \\ $\delta=5\%$ \\ $\ \, \delta=10\%$ } &
        $\makecell{ 3.61\times10^{-14} \\ 4.63\times10^{-14} \\ 5.91\times10^{-14}}$ &
        $\makecell{ 5.47\times10^{-14} \\ 6.94\times10^{-14} \\ 8.87\times10^{-14}}$ &
        $\makecell{ 4.53\times10^{-14} \\ 5.51\times10^{-14} \\ 6.94\times10^{-14}}$ \\
        \hline
    \end{tabular}
\end{table}
%
\begin{table}[h]
    \centering \caption{The same as in Tab.~\ref{tab:excl_750}, but for
        the energy $\sqrt{s} = 3000$ GeV and different values of the
        integrated luminosities. \bigskip} \label{tab:excl_1500}
    \begin{tabular}{||c||c|c|c|c||}
        \hline
        $\lambda_e$ & & 0 & $-0.8$ & 0.8 \\
        \hline
        \quad $L$, fb$^{-1}$ & & 5000 & 4000 & 1000 \\
        \hline
        \makecell{$|g_1|$, GeV$^{-4}$ \\ $(g_2=0)$ } & \makecell{$\delta=0\%$ \\ $\delta=5\%$ \\ $\ \, \delta=10\%$ } &
        $\makecell{ 5.98\times10^{-15} \\ 1.33\times10^{-14} \\ 1.85\times10^{-14}}$ &
        $\makecell{ 7.14\times10^{-15} \\ 1.73\times10^{-14} \\ 2.39\times10^{-14}}$ &
        $\makecell{ 5.13\times10^{-15} \\ 7.79\times10^{-15} \\ 1.04\times10^{-14}}$ \\
        \hline
        \makecell{$|g_2|$, GeV$^{-4}$ \\ $(g_1=0)$} & \makecell{$\delta=0\%$ \\ $\delta=5\%$ \\ $\ \, \delta=10\%$ } &
        $\makecell{ 5.18\times10^{-15} \\ 1.16\times10^{-14} \\ 1.62\times10^{-14}}$ &
        $\makecell{ 6.62\times10^{-15} \\ 1.60\times10^{-14} \\ 2.21\times10^{-14}}$ &
        $\makecell{ 5.19\times10^{-15} \\ 7.87\times10^{-15} \\ 1.05\times10^{-14}}$ \\
        \hline
    \end{tabular}
\end{table}

Recently, the bounds on the anomalous quartic couplings for the
vertex $\gamma\gamma\gamma Z$ were obtained via $\gamma Z$
production with intact protons in the forward region at the LHC
\cite{Baldenegro:2017}. To examine this process, the effective
Lagrangian \eqref{Lagrangian_B} was used with the anomalous
couplings $\zeta, \tilde{\zeta}$. Both for integrated luminosity 300
fb$^{-1}$ and high luminosity 3000 fb$^{-1}$ sensitivities were
found to be similar, $\zeta, \tilde{\zeta} \sim 1 \times 10^{-13}$
at the 95\% C.L. Taking into account the relations between couplings
$\zeta, \tilde{\zeta}$ and our couplings $g_1, g_2$,
\eqref{couplings_G-R_B}, we expect that the sensitivities of $g_1,
g_2 \sim 8\times 10^{-13}$ can be reached at the LHC (HL-LHC). These
values should be compared with the CLIC bounds in
Tabs.~\ref{tab:excl_750} and \ref{tab:excl_1500}. Note that the
expected sensitivity from the $Z\rightarrow\gamma\gamma\gamma$ decay
search at the LHC \cite{ATLAS:Z_decay} is approximately three orders
of magnitude smaller than that obtained in \cite{Baldenegro:2017}.

\section{Unitarity constraints on anomalous quartic couplings} %
\label{sec:unit_const}

The anomalous contribution to the total cross section rises as
$s^3$. Thus, the contribution of the effective operators in
\eqref{Lagrangian} may lead to unitarity violation at high energies.
That is why we need to study bounds imposed by partial-wave
unitarity. The partial-wave expansion of the helicity amplitude in
the center-of-mass system was derived in \cite{Jacob:2000} and used
in a number of papers \cite{Gounaris:1994}. It looks like
\begin{align}\label{helicity_ampl_expansion}
M_{\lambda_1\lambda_2\lambda_3\lambda_4}(s, \theta, \varphi) &=
16\pi \sum_J (2J + 1) \sqrt{(1 + \delta_{\lambda_1\lambda_2})(1 +
\delta_{\lambda_3\lambda_4})}
\nonumber \\
&\times \,e^{i(\lambda - \mu)\phi} \,d^J_{\lambda\mu}(\theta)
\,T^J_{\lambda_1\lambda_2\lambda_3\lambda_4}(s) \;,
\end{align}
where $\lambda = \lambda_1 - \lambda_2$, $\mu = \lambda_3 -
\lambda_4$, $\theta(\phi)$ is the polar (azimuth) scattering angle,
and $d^J_{\lambda\mu}(\theta)$ is the Wigner (small) $d$-function
\cite{Wigner}. Relevant formulas for the $d$-functions are given in
Appendix~C. In our case $\lambda, \mu$ are even numbers, $\lambda,
\,\mu = 0, \pm 2$ (see below). If we choose the plane $(x - z)$ as a
scattering plane, then $\phi = 0$ in
\eqref{helicity_ampl_expansion}. Parity conservation means that
\begin{equation}\label{parity_conservation}
T^J_{\lambda_1\lambda_2\lambda_3\lambda_4}(s) = (-1)^{\lambda_1 -
\lambda_2 - \lambda_3 + \lambda_4}
\,T^J_{-\lambda_1-\lambda_2-\lambda_3-\lambda_4}(s) \;.
\end{equation}
Partial-wave unitarity in the limit $s \gg (m_1 + m_2)^2$ requires
that
\begin{equation}\label{parity_wave_unitarity}
\left| T^J_{\lambda_1\lambda_2\lambda_3\lambda_4}(s) \right| \leq 1
\;.
\end{equation}
Using orthogonality of the $d$-functions
\eqref{Wigner_func_orthogonality}, we find the partial-wave
amplitude
\begin{align}\label{parity wave_func}
T^J_{\lambda_1\lambda_2\lambda_3\lambda_4}(s) &= \frac{1}{32\pi}
\frac{1}{\sqrt{(1 + \delta_{\lambda_1\lambda_2})(1 +
\delta_{\lambda_3\lambda_4})}} \int\limits_{-1}^1 \!\!
M_{\lambda_1\lambda_2\lambda_3\lambda_4}(s, z)
\,d^{J}_{\lambda\mu}(z) \,dz \;.
\end{align}
Here and in what follows, $z = \cos\theta$. Note that
$M_{\lambda_1\lambda_2\lambda_3\lambda_4} = g_1
M^{(1)}_{\lambda_1\lambda_2\lambda_3\lambda_4} + g_2
M^{(2)}_{\lambda_1\lambda_2\lambda_3\lambda_4}$, and the helicity
amplitudes $M^{(1,2)}_{\lambda_1\lambda_2\lambda_3\lambda_4}$ are
given in Appendix~B.

\subsection{Unitarity bounds on coupling $\bs{g_1}$ ($\bs{g_2 = 0}$)} %

To obtain a unitarity bound on the coupling $g_1$, we put $g_2 = 0$.
Let us note that due to eq.~\eqref{parity_conservation}, it is
sufficient to examine the helicity amplitudes with $\lambda_1 = +1$
only. Moreover, it is enough to consider four amplitudes,
$M^{(1)}_{++++}$, $M^{(1)}_{+-+-}$, $M^{(1)}_{+--+}$, and
$M^{(1)}_{++--}$, since the rest are suppressed by small factor
$m_Z/\sqrt{s}$ or zero. \\

\textbf{1.} $\lambda_1 = \lambda_2 = \lambda_3 = \lambda_4 = 1$,
then $\lambda = \mu = 0$. The helicity amplitude is given by the
first of equations \eqref{independent_ampl_M1},
\begin{equation}\label{M1_++++}
M_{++++}(s,z) = g_1 M^{(1)}_{++++}(s) = -\frac{g_1 }{4} \,s(s -
m_Z^2) \;.
\end{equation}
Using eqs.~\eqref{d-function_00}-\eqref{integral_Legendre}, we find
that the partial-wave amplitude with $J=0$ is the only non-zero
amplitude,
\begin{equation}\label{T1_++++}
T^0_{++++}(s) = - \frac{g_1}{128\pi} \,s(s - m_Z^2)
\int\limits_{-1}^1 d^{\,0}_{00}(z) \,dz = - \frac{g_1}{64\pi} \,s(s
- m_Z^2) \;.
\end{equation}
Correspondingly, we obtain from \eqref{parity_wave_unitarity},
\eqref{T1_++++}
\begin{equation}\label{T1++++_unitarity}
|g_1| \leq 64\pi [s(s - m_Z^2)]^{-1} \;.
\end{equation}

\textbf{2.} $\lambda_1 = -\lambda_2 = \lambda_3 = -\lambda_4 = 1$,
then $\lambda = \mu = 2$. According to \eqref{independent_ampl_M1},
\begin{equation}\label{M1_+-+-}
M_{+-+-}(s,z) = g_1 M^{(1)}_{+-+-}(s,z) = -\frac{g_1}{4}
\,\frac{s^3}{s - m_Z^2} \left( \frac{1 + z}{2}\right)^{\!2} .
\end{equation}
It follows from \eqref{parity wave_func}, \eqref{d-function_22} that
\begin{align}\label{T1+-+-_J}
T^J_{+-+-}(s) &= -\frac{g_1}{128\pi} \,\frac{s^3}{s - m_Z^2}
\int\limits_{-1}^1 \left( \frac{1 + z}{2}\right)^{\!2}
\,d^{J}_{22}(z) \,dz = -\frac{g_1}{128\pi} \,\frac{s^3}{s - m_Z^2}
\nonumber \\
&\times \int\limits_{-1}^1 \left( \frac{1 + z}{2}\right)^{\!4}
{}_2F_1 \left( 2-J, J+3; 1; \frac{1-z}{2} \right) dz \;.
\end{align}
Let $(1 - z)/2 = x$, then $(1 + z)/2 = 1 - x$, and we find
\begin{align}\label{}
T^J_{+-+-}(s) &= -\frac{1}{64\pi} \,\frac{g_1 s^3}{s - m_Z^2}
\int\limits_0^1 (1 - x)^4 \,{}_2F_1( 2-J, J+3; 1; x) \,dx
\nonumber \\
&= -\frac{3g_1}{8\pi\Gamma(4+J)\Gamma(3-J)} \frac{s^3}{s - m_Z^2}
\;,
\end{align}
where we used formulas 2.21.1.5 and 2.21.1.6 in \cite{Prudnikov_v3}.
Thus, only three partial-waves amplitudes, $T^0_{+-+-}(s)$,
$T^1_{+-+-}(s)$, and $T^2_{+-+-}(s)$, are non-zero. The most
important for us is $T^0_{+-+-}(s)$, since it results in the
strongest constraint on the coupling $g_1$,
\begin{equation}\label{T1+-+-_unitarity}
|g_1| \leq 32\,\pi (s - m_Z^2) s^{-3}\;.
\end{equation}

\textbf{3.} $\lambda_1 = -\lambda_2 = -\lambda_3 = \lambda_4 = 1$,
then $\lambda = 2, \,\mu = -2$, and we have
\begin{equation}\label{M1_+--+}
M_{+--+}(s,z) = g_1 M^{(1)}_{+--+}(s,z) = -\frac{g_1}{4}
\,\frac{s^3}{s - m_Z^2} \left( \frac{1 - z}{2}\right)^{\!2} .
\end{equation}
Using eq.~\eqref{d-function_2-2} and first relation in
\eqref{d-function_symmetry}, after substitutions $(1 + z)/2 = x$,
$(1 - z)/2 = 1 - x$, we reduce this case to the previous one. As a
result, we come again to the upper bound \eqref{T1+-+-_unitarity}.

\textbf{4.} $\lambda_1 = \lambda_2 = -\lambda_3 = -\lambda_4 = 1$,
then $\lambda = \mu = 0$, and
\begin{equation}\label{M1_++--}
M_{++--}(s,z) = g_1 M^{(1)}_{++--}(s,z) = -\frac{g_1}{8}
\,\frac{s^3}{s - m_Z^2} (3 + z^2) \;.
\end{equation}
Only two partial-waves amplitudes, $T^0_{++--}(s)$ and
$T^2_{++--}(s)$, are non-zero,
\begin{equation}\label{T1_++--}
T^0_{++--}(s) =  -\frac{5\,g_1}{192\pi} \,\frac{s^3}{s - m_Z^2} \;,
\quad T^2_{++--}(s) =  -\frac{g_1}{960\pi} \,\frac{s^3}{s - m_Z^2}
\;.
\end{equation}
The strongest bound on $g_1$ comes from unitarity constraint on
$T^0_{++--}(s)$,
\begin{equation}\label{T1++--_unitarity}
|g_1| \leq \frac{192\,\pi}{5} (s - m_Z^2) s^{-3} \;.
\end{equation}

\subsection{Unitarity bounds on coupling $\bs{g_2}$ ($\bs{g_1 = 0}$)} %

To derive a unitarity bound on the coupling $g_2$, we take $g_1 =
0$. It is sufficient to consider three amplitudes, $M^{(2)}_{++++}$,
$M^{(2)}_{+-+-}$, and $M^{(2)}_{+--+}$. The rest are suppressed by
small factor $m_Z/\sqrt{s}$ or zero. \\

\textbf{1.} $\lambda_1 = \lambda_2 = \lambda_3 = \lambda_4 = 1$,
then $\lambda = \mu = 0$. The helicity amplitude is given by the
first of equations \eqref{independent_ampl_M2}
\begin{equation}\label{M2_++++}
M_{++++}(s,z) = g_2 M^{(2)}_{++++}(s) = \frac{g_2}{2} \,s(s - m_Z^2)
\;.
\end{equation}
As a result, we get
\begin{equation}\label{T2++++_unitarity}
|g_2| \leq 32\pi [s(s - m_Z^2)]^{-1} \;.
\end{equation}

\textbf{2.} $\lambda_1 = -\lambda_2 = \lambda_3 = -\lambda_4 = 1$,
then $\lambda = \mu = 2$, and we find from
\eqref{independent_ampl_M2}
\begin{equation}\label{M2_+-+-}
M_{+-+-}(s,z) = g_2 M^{(2)}_{+-+-}(s,z) = \frac{g_2}{2}
\,\frac{s^3}{s - m_Z^2} \left( \frac{1 + z}{2}\right)^{\!2} .
\end{equation}
We follow the derivation of eq.~\eqref{T1+-+-_unitarity} and come to
the inequality
\begin{equation}\label{T2+-+-_unitarity}
|g_2|  \leq 16\pi (s - m_Z^2) s^{-3} \;.
\end{equation}

\textbf{3.} $\lambda_1 = -\lambda_2 = -\lambda_3 = \lambda_4 = 1$,
then $\lambda = 2, \,\mu = -2$, and we obtain
\begin{equation}\label{M2_+--+}
M_{+--+}(s,z) = g_2 M^{(2)}_{+--+}(s,z) = \frac{g_2}{2}
\,\frac{s^3}{s - m_Z^2} \left( \frac{1 - z}{2}\right)^{\!2} .
\end{equation}
Using the first relation in \eqref{d-function_symmetry} and
eq.~\eqref{d-function_2-2}, we can reduce this case  to the previous
one to get eq.~\eqref{T2+-+-_unitarity}.

\textbf{4.} $\lambda_1 = \lambda_2 = -\lambda_3 = -\lambda_4 = 1$.
The helicity amplitude $M^{(2)}_{++--}(s,z) = 0$.

\subsection{Unitarity bounds on couplings $\bs{g_1}$ and $\bs{g_2}$} %

Now we consider a general case with $g_1, g_2 \neq 0$. Note that
$M^{(2)}_{++++} = - 2M^{(1)}_{++++}$, $M^{(2)}_{+-+-} =
-2M^{(1)}_{+-+-}$, and $M^{(2)}_{+--+} = - 2M^{(1)}_{+--+}$.
Correspondingly, $M_{++++} = (g_1 - 2g_2)M^{(1)}_{++++}$, etc. From
formulas derived in two previous subsections we immediately get the
following bound on a linear combination of $g_1$ and $g_2$,
\begin{equation}\label{g1_g2_unitarity}
|g_1 - 2g_2| \leq 32\,\pi (s - m_Z^2) s^{-3} \;.
\end{equation}
Let us underline that $M_{++--}(s,z) = g_1 M^{(1)}_{++--}(s,z)$. It
means that inequality \eqref{T1++--_unitarity} holds for a general
case ($g_1, g_2 \neq 0$). It enables us to obtain constraints
separately on each coupling. If the couplings $g_1$, $g_2$ have the
same sign, then
\begin{equation}\label{g1_g2_same_signs}
|g_1| \leq  \frac{192\,\pi}{5} (s - m_Z^2) s^{-3}  \;, \quad |g_2|
\leq \frac{176\,\pi}{5}(s - m_Z^2) s^{-3}  \;.
\end{equation}
If the signs of the couplings $g_1$, $g_2$ are opposite, we obtain
\begin{equation}\label{g1_g2_opposite_signs}
|g_1| \leq  32\,\pi \,(s - m_Z^2) s^{-3} \;, \quad |g_2| \leq
16\,\pi \,(s - m_Z^2) s^{-3}  \;.
\end{equation}
The bounds on the couplings $g_1, g_2$ along with their numerical
values are collected in Tab.~\ref{tab:unit_lim}. We have taken into
account that $m_Z^2/s \ll 1$ for the CLIC energies.

{\setlength{\extrarowheight}{4pt}
\begin{table}[h]
\centering \caption{Unitarity constraints on the anomalous couplings
when just one coupling is non-zero (second and third columns), and
when both couplings are non-vanishing (fourth and fifth columns for
the couplings of the same sign, sixth and seventh columns for the
couplings of opposite signs). The numerical values of the bounds are
given for the collision energy $\sqrt{s} = 1500(3000)$ GeV.
\bigskip} \label{tab:unit_lim}
\begin{tabular}{||c||c|c||c|c||c|c||}
  \hline
\multicolumn{1}{||c||} {} & \multicolumn{2}{c||} {1 operator ($g_2 =
0$ or $g_1 = 0$)} & \multicolumn{2}{c||} {2 operators ($g_1 g_2 >
0$)} & \multicolumn{2}{c||} {2 operators ($g_1 g_2 < 0$)}  \\
  \hline
  $g_1$ & $32\pi s^{-2}$ & 20(1.2) TeV$^{-4}$ & $\frac{192}{5}s^{-2}$
  & 24(1.5) TeV$^{-4}$ & $32\pi s^{-2}$ & 20(1.2) TeV$^{-4}$ \\ [2pt]
  \hline
  $g_2$  & $16\pi s^{-2}$ & 10(0.6) TeV$^{-4}$ & $\frac{176}{5}s^{-2}$
  & 22(1.4) TeV$^{-4}$ & $16\pi s^{-2}$ & 10(0.6) TeV$^{-4}$ \\ [2pt]
  \hline
\end{tabular}
\end{table}
}

To summarize, in spite of the fact that the anomalous contribution
to the total cross section is proportional to $s^3$, the unitarity
is not violated in the region of the anomalous QGCs presented in
Tabs.~\ref{tab:excl_750}, \ref{tab:excl_1500}.

\section{Conclusions} %
\label{sec:concl}

In the present paper, the CLIC discovery potential for exclusive
$\gamma Z$ production in the scattering of the Compton backscattered
photons at the $e^+e^-$ collision energies 1500 GeV and 3000 GeV is
studied. We have shown that such a process provides an opportunity
of searching for the anomalous quartic neutral gauge couplings for
the $\gamma\gamma\gamma Z$ vertex at the CLIC. Both unpolarized and
polarized initial electron beams are examined. To describe the
anomalous quartic gauge couplings we used the effective Lagrangian
which conserves gauge invariance. Although quartic gauge couplings
are already induced at the dimension-six level, we considered the
effective Lagrangian with CP conserving dimension-eight operators
without contributing to anomalous trilinear gauge interactions.

We have derived the explicit expressions for the anomalous
contributions to the helicity amplitudes of the process
$\gamma\gamma \rightarrow \gamma Z$. After that the differential and
total cross sections are calculated depending on $m_{Z\gamma}$, the
invariant mass of the $\gamma Z$ system. It is shown that the
anomalous contribution dominates both the interference and SM cross
sections. Moreover, the ratio of the total cross section to the SM
one grows with the increase of $m_{Z\gamma}$, being more
approximately one order of magnitude at large $m_{\gamma Z}$.

It enabled us to obtain the exclusion regions for the anomalous
couplings with the systematic errors of 0\%, 5\%, and 10\%. We have
considered the $Z$ boson decay into leptons (electron and muons).
For both couplings, $g_{1,2}$, the best bounds are equal to
approximately $4.4 \times 10^{-14}$ GeV$^{-4}$ and $5.1 \times
10^{-15}$ GeV$^{-4}$, for the $e^+e^-$ energies 1500 GeV and 3000
GeV, respectively. They are achieved when electron beam helicity is
equal to 0.8. We have checked that the unitarity is not violated in
the region of the couplings considered in the paper. Our best bound
on the anomalous couplings for the collision energy 3000 GeV is
roughly two orders of magnitude stronger than the limits which can
be reached at the LHC and HL-LHC. This points to a great potential
of the CLIC and other future leptonic colliders to probe the
anomalous $\gamma\gamma\gamma Z$ couplings.



\setcounter{equation}{0}
\renewcommand{\theequation}{A.\arabic{equation}}

\section*{Appendix A}
\label{app:A}

Here we present explicit expressions for components of the
polarization tensor \eqref{polarization_tensor}. They are the
following
\begin{align}\label{P1.1}
P_{\mu\nu\rho\alpha}^{(1.1)} &= (p_1\cdot p_2)[ (p_1\cdot p_3) +
(p_2\cdot p_3)] g_{\mu\nu}g_{\rho\alpha} + (p_1\cdot p_3)[ (p_1\cdot
p_2) + (p_2\cdot p_3) ] g_{\mu\rho}g_{\nu\alpha}
\nonumber \\
&+ (p_2\cdot p_3)[ (p_1\cdot p_2) + (p_1\cdot p_3) ]
g_{\nu\rho}g_{\mu\alpha} \;,
\end{align}
\begin{align}\label{P1.2}
P_{\mu\nu\rho\alpha}^{(1.2)} &= - \{ [ (p_1\cdot p_2) + (p_1\cdot
p_3)] p_{3\nu} p_{2\rho}g_{\mu\alpha} + [ (p_1\cdot p_2) + (p_2\cdot
p_3) ] p_{3\mu}p_{1\rho} g_{\nu\alpha}
\nonumber \\
&\quad + [ (p_1\cdot p_3) + (p_2\cdot p_3) ]
p_{2\mu}p_{1\nu}g_{\rho\alpha} \} \;,
\end{align}
\begin{align}\label{P1.3}
P_{\mu\nu\rho\alpha}^{(1.3)} &= -[ (p_1\cdot p_2)( p_{1\rho} +
p_{2\rho} )p_{3\alpha}g_{\mu\nu} + (p_1\cdot p_3)( p_{1\nu} +
p_{3\nu} )p_{2\alpha} g_{\mu\rho}
\nonumber \\
&\quad + (p_2\cdot p_3)( p_{2\mu} + p_{3\mu} )p_{1\alpha}g_{\nu\rho}
] \;,
\end{align}
\begin{align}\label{P1.4}
P_{\mu\nu\rho\alpha}^{(1.4)} &= p_{2\mu}p_{1\nu} ( p_{1\rho} +
p_{2\rho} ) p_{3\alpha} + p_{3\mu} ( p_{1\nu} + p_{3\nu} ) p_{1\rho}
p_{2\alpha}
\nonumber \\
&+ p_{3\nu} p_{2\rho}( p_{2\mu} + p_{3\mu} ) p_{1\alpha} \;,
\end{align}
and
\begin{align}\label{P2.1}
P_{\mu\nu\rho\alpha}^{(2.1)} = &- 2[(p_1\cdot p_3)(p_2\cdot p_3)
g_{\mu\nu}g_{\rho\alpha} + (p_1\cdot p_2)(p_2\cdot p_3)
g_{\mu\rho}g_{\nu\alpha}
\nonumber \\
&+ (p_1\cdot p_2)(p_1\cdot p_3) g_{\nu\rho} g_{\mu\alpha}] \;,
\end{align}
\begin{align}\label{P2.2}
P_{\mu\nu\rho\alpha}^{(2.2)} &= (p_1\cdot
p_2)[p_{3\mu}p_{3\alpha}g_{\nu\rho} +
p_{3\nu}p_{3\alpha}g_{\mu\rho}] + (p_1\cdot
p_3)[p_{2\mu}p_{2\alpha}g_{\nu\rho} +
p_{2\rho}p_{2\alpha}g_{\mu\nu}]
\nonumber \\
&+ (p_2\cdot p_3)[p_{1\nu}p_{1\alpha}g_{\mu\rho} +
p_{1\rho}p_{1\alpha}g_{\mu\nu}] \;,
\end{align}
\begin{equation}\label{P2.3}
P_{\mu\nu\rho\alpha}^{(2.3)} = - 2[(p_1\cdot p_2)p_{3\mu}p_{3\nu}
g_{\rho\alpha} + (p_1\cdot p_3)p_{2\mu}p_{2\rho} g_{\nu\alpha} +
(p_2\cdot p_3)p_{1\nu}p_{1\rho} g_{\mu\alpha}] \;,
\end{equation}
\begin{align}\label{P2.4}
P_{\mu\nu\rho\alpha}^{(2.4)} &= (p_1\cdot
p_2)[p_{3\mu}p_{1\alpha}g_{\nu\rho} +
p_{3\nu}p_{2\alpha}g_{\mu\rho}] + (p_1\cdot
p_3)[p_{2\mu}p_{1\alpha}g_{\nu\rho} +
p_{2\rho}p_{3\alpha}g_{\mu\nu}]
\nonumber \\
&+ (p_2\cdot p_3)[p_{1\nu}p_{2\alpha}g_{\mu\rho} +
p_{1\rho}p_{3\alpha}g_{\mu\nu}] \;,
\end{align}
\begin{align}\label{P2.5}
P_{\mu\nu\rho\alpha}^{(2.5)} &= 2\{ (p_1\cdot
p_2)[p_{3\mu}p_{2\rho}g_{\nu\alpha} +
p_{3\nu}p_{1\rho}g_{\mu\alpha}] + (p_1\cdot
p_3)[p_{2\mu}p_{3\nu}g_{\rho\alpha} +
p_{2\rho}p_{1\nu}g_{\mu\alpha}]
\nonumber \\
&+ (p_2\cdot p_3)[p_{3\mu}p_{1\nu}g_{\rho\alpha} +
p_{2\mu}p_{1\rho}g_{\nu\alpha}] \} \;,
\end{align}
\begin{align}\label{P2.6}
P_{\mu\nu\rho\alpha}^{(2.6)} = &-\{ (p_1\cdot
p_2)[p_{3\mu}p_{2\alpha}g_{\nu\rho} +
p_{3\nu}p_{1\alpha}g_{\mu\rho}] + (p_1\cdot
p_3)[p_{2\mu}p_{3\alpha}g_{\nu\rho} +
p_{2\rho}p_{1\alpha}g_{\mu\nu}]
\nonumber \\
&+ (p_2\cdot p_3)[p_{1\nu}p_{3\alpha}g_{\mu\rho} +
p_{1\rho}p_{2\alpha}g_{\mu\nu}] \} \;,
\end{align}
\begin{align}\label{P2.7}
P_{\mu\nu\rho\alpha}^{(2.7)} = - ( p_{2\mu}p_{3\nu}p_{1\rho} +
p_{3\mu}p_{1\nu}p_{2\rho} ) (p_{1\alpha} + p_{2\alpha} +
p_{3\alpha}) \;.
\end{align}
Note that the last tensor does not contribute to the matrix element
\eqref{matrix_element}, since it is proportional to $p_{4\alpha}$.
One can directly check that
\begin{align}\label{polarization_tensor_gauge_inv}
p_1^\mu \sum_{i=1}^4 P_{\mu\nu\rho\alpha}^{(1.i)} &= p_2^\nu
\sum_{i=1}^4 P_{\mu\nu\rho\alpha}^{(1.i)} = p_3^\rho \sum_{i=1}^4
P_{\mu\nu\rho\alpha}^{(1.i)} = 0 \;,
\nonumber \\
p_1^\mu \sum_{i=1}^7 P_{\mu\nu\rho\alpha}^{(2.i)} &= p_2^\nu
\sum_{i=1}^7 P_{\mu\nu\rho\alpha}^{(2.i)} = p_3^\rho \sum_{i=1}^7
P_{\mu\nu\rho\alpha}^{(2.i)} = 0 \;.
\end{align}



\setcounter{equation}{0}
\renewcommand{\theequation}{B.\arabic{equation}}

\section*{Appendix B}
\label{app:B}

In accordance with eq.~\eqref{polarization_tensor}, any anomalous
helicity amplitude is the sum of two terms,
\begin{equation}\label{helicity_ampl_sum}
M_{\lambda_1\lambda_2\lambda_3\lambda_4} = g_1
M_{\lambda_1\lambda_2\lambda_3\lambda_4}^{(1)} + g_2
M_{\lambda_1\lambda_2\lambda_3\lambda_4}^{(2)} \;.
\end{equation}
There are $2^3\times3 = 24$ helicity amplitudes
$M^{(1)}_{\lambda_1\lambda_2\lambda_3\lambda_4}$ and,
correspondingly, 24 amplitudes
$M^{(2)}_{\lambda_1\lambda_2\lambda_3\lambda_4}$ for the process
\eqref{process}. Bose-Einstein statistics and parity invariance
demand that there exist nine independent helicity amplitudes
$M^{(1)}_{\lambda_1\lambda_2\lambda_3\lambda_4}$ with $\lambda_1 =
+1$, six for transverse $Z$ and three for longitudinal $Z$. Our
calculations resulted in the following helicity amplitudes
$M^{(1)}_{\lambda_1\lambda_2\lambda_3\lambda_4}$ with $\lambda_1 =
+1$
\begin{align}\label{independent_ampl_M1}
M^{(1)}_{++++}(s,t,u) &= \frac{1}{4} s(t + u) \;,
\nonumber \\
M^{(1)}_{+++-}(s,t,u) &= 0 \;,
\nonumber \\
M^{(1)}_{++-+}(s,t,u) &= \frac{1}{2} \frac{tu}{t + u} m_Z^2 \;,
\nonumber \\
M^{(1)}_{++--}(s,t,u) &= \frac{1}{2}\frac{s(t^2 + tu + u^2)}{t + u}
\;,
\nonumber \\
M^{(1)}_{+-++}(s,t,u) &= \frac{1}{4} \frac{tu}{t + u} m_Z^2 \;,
\nonumber \\
M^{(1)}_{+-+-}(s,t,u) &= \frac{1}{4}\frac{su^2}{t + u} \;,
\nonumber \\
M^{(1)}_{+++0}(s,t,u) &= 0 \;,
\nonumber \\
M^{(1)}_{++-0}(s,t,u) &= \frac{i}{2\sqrt{2}}\sqrt{stu} \,\frac{t -
u}{t + u} \,m_Z \;,
\nonumber \\
M^{(1)}_{+-+0}(s,t,u) &=  -\frac{i}{2\sqrt{2}}\frac{u\sqrt{stu}}{t +
u} \,m_Z \;.
\end{align}
Three more amplitudes $M^{(1)}_{+\lambda_2\lambda_3\lambda_4}$ can
be obtained by exchanging Mandelstam variables $t$ and
$u$~\cite{Gounaris:1999_1,Glover:1993},
\begin{align}\label{dependent_ampl_M1}
M^{(1)}_{+--+}(s,t,u) &= M^{(1)}_{+-+-}(s,u,t) =
\frac{1}{4}\frac{st^2}{t + u}\;,
\nonumber \\
M^{(1)}_{+---}(s,t,u) &= M^{(1)}_{+-++}(s,u,t) = \frac{1}{4}
\frac{tu}{(t + u)} m_Z^2 \;,
\nonumber \\
M^{(1)}_{+--0}(s,t,u) &= M^{(1)}_{+-+0}(s,u,t) =
-\frac{i}{2\sqrt{2}}\frac{t\sqrt{stu}}{t + u}\,m_Z \;.
\end{align}

Nine independent helicity amplitudes
$M^{(2)}_{\lambda_1\lambda_2\lambda_3\lambda_4}$ with $\lambda_1 =
+1$ are
\begin{align}\label{independent_ampl_M2}
M^{(2)}_{++++}(s,t,u) &= -\frac{1}{2} s(t + u) \;,
\nonumber \\
M^{(2)}_{+++-}(s,t,u) &= 0 \;,
\nonumber \\
M^{(2)}_{++-+}(s,t,u) &= 0 \;,
\nonumber \\
M^{(2)}_{++--}(s,t,u) &= 0 \;,
\nonumber \\
M^{(2)}_{+-++}(s,t,u) &= - \frac{1}{2} \frac{tu}{t + u} m_Z^2 \;,
\nonumber \\
M^{(2)}_{+-+-}(s,t,u) &= -\frac{1}{2}\frac{su^2}{t + u} \;,
\nonumber \\
M^{(2)}_{+++0}(s,t,u) &= 0 \;,
\nonumber \\
M^{(2)}_{++-0}(s,t,u) &= 0 \;,
\nonumber \\
M^{(2)}_{+-+0}(s,t,u) &= \frac{i}{\sqrt{2}}\frac{u\sqrt{stu}}{t +
u}m_Z \;.
\end{align}
The other three  helicity amplitudes
$M^{(2)}_{+\lambda_2\lambda_3\lambda_4}$ are given by
\begin{align}\label{dependent_ampl_M2}
M^{(2)}_{+--+}(s,t,u) &= M^{(2)}_{+-+-}(s,u,t) =
-\frac{1}{4}\frac{st^2}{t + u} \;,
\nonumber \\
M^{(2)}_{+---}(s,t,u) &= M^{(2)}_{+-++}(s,u,t) = - \frac{1}{2}
\frac{tu}{t + u} m_Z^2 \;,
\nonumber \\
M^{(2)}_{+--0}(s,t,u) &= M^{(2)}_{+-+0}(s,u,t) =
\frac{i}{\sqrt{2}}\frac{t\sqrt{stu}}{t + u}m_Z \;.
\end{align}
Note that all amplitudes $M^{(1,2)}_{\lambda_1\lambda_2\lambda_30}$
are equal to zero in the limit $m_Z = 0$.

The amplitudes with $\lambda_1 = -1$ can be obtained from
constraints imposed by parity
invariance~\cite{Gounaris:1999_1,Glover:1993},
\begin{equation}\label{parity_relations}
M^{(1,2)}_{-\lambda_2\lambda_3\lambda_4}(s,t,u) = (-1)^{1 -
\lambda_4} M^{(1,2)}_{+-\lambda_2-\lambda_3-\lambda_4}(s,t,u) \;.
\end{equation}
Note that we have directly calculated all 48 helicity amplitudes
using eq.~\eqref{matrix_element}. Our calculations show that
relations \eqref{dependent_ampl_M1}, \eqref{dependent_ampl_M2}, and
\eqref{parity_relations} really hold.



\setcounter{equation}{0}
\renewcommand{\theequation}{C.\arabic{equation}}

\section*{Appendix C}
\label{app:C}

Wigner's $d$-functions \cite{Wigner} are related to the Jacobi
polynomials $P^{(\alpha, \,\beta)}_n(z)$ with nonnegative $\alpha,
\beta$ \cite{Varshalovich},
\begin{align}\label{d-function}
d^J_{\lambda\mu}(z) &= \left[ \frac{(J + \lambda)!(J - \lambda)!}{(J
+ \mu)!(J - \mu)!}\right]^{1/2} \left( \frac{1 -
z}{2}\right)^{(\lambda - \mu)/2} \left( \frac{1 +
z}{2}\right)^{(\lambda + \mu)/2}
\nonumber \\
&\times P^{(\lambda-\mu, \,\lambda+\mu)}_{J - \lambda}(z) \;,
\end{align}
where $z = \cos\theta$. The $d$-functions obey the orthogonality
condition \cite{Varshalovich}
\begin{equation}\label{Wigner_func_orthogonality}
\int\limits_{-1}^1 d^{J}_{\lambda\lambda'}(z)
\,d^{J'}_{\lambda\lambda'}(z) \,dz = \frac{2}{2J + 1} \,\delta_{JJ'}
\;.
\end{equation}
In its turn, the Jacobi polynomial is related to the hypergeometric
function \cite{Varshalovich},
\begin{equation}\label{Jacobi_polynomial}
P^{(\rho, \,\sigma)}_n(z) = \frac{\Gamma(\rho + 1 + n)}{\Gamma(\rho
+ 1) \,n!} \,{}_2F_1 \!\left( -n, \rho+\sigma + n + 1; \rho +1;
\frac{1-z}{2} \right) .
\end{equation}
Note that $P^{(\alpha, \,\beta)}_n(-z) = (-1)^n P^{(\beta,
\,\alpha)}_n(z)$, and, correspondingly,
\begin{equation}\label{d-function_symmetry}
d^J_{\lambda\mu}(-z) = (-1)^{J - \lambda}d^J_{\mu -\lambda}(z) \;,
\quad d^J_{\lambda\mu}(z) = (-1)^{\lambda - \mu}d^J_{-\lambda
-\mu}(z) \;.
\end{equation}
In particular, we get
\begin{align}
d^J_{22}(z) &= \left( \frac{1 + z}{2}\right)^{\!2} {}_2F_1 \!\left(
2-J, J+3; 1; \frac{1-z}{2} \right) , \label{d-function_22} \\
d^J_{2-2}(z) &= (-1)^J \left( \frac{1 - z}{2}\right)^{\!2} {}_2F_1
\!\left( 2-J, J+3; 1; \frac{1+z}{2} \right) . \label{d-function_2-2}
\end{align}
In the simplest case, $\lambda = \mu = 0$, we find
\begin{align}\label{d-function_00}
d^J_{00}(z) &= P_J(z) \;,
\end{align}
where $P_J(z)$ being the Legendre polynomial. Using table integral
7.231.1 in \cite{Gradshteyn}, we derive the following formula
\begin{equation}\label{integral_Legendre}
\int\limits_{-1}^1 z^m P_J(z) \,dz = \frac{1}{2} \left[ 1 + (-1)^J
\right] (-1)^{J/2} \frac{\Gamma \!\left( \frac{J-m}{2} \right)
\Gamma \!\left( \frac{1+m}{2}\right)}{{\Gamma \!\left(
-\frac{m}{2}\right) \Gamma \!\left( \frac{J+m+3}{2}\right)}} \;,
\end{equation}
with integer $J$ and even number $m \geq 0$. To obtain unitarity
constraints on the anomalous couplings, we need integral
\eqref{integral_Legendre} with $m=0, \,2$.




\end{document}